\documentclass[preprint,12pt]{elsarticle}

\usepackage{amssymb}
\usepackage{amsmath}

\usepackage{siunitx}
\usepackage{booktabs}
\usepackage{tikz}

\newcommand{\bigo}[1]{\mathcal{O}\left( #1 \right)}
\usepackage{subcaption}
\usepackage{hyperref}
\hypersetup{colorlinks,allcolors=black}

\makeatother

\journal{Chaos, Solitons \& Fractals}

\begin{document}

\begin{frontmatter}



\title{Equivalence of stationary dynamical solutions in a directed chain and a Delay Differential Equation of neuroscientific relevance}


\author[unibo,infnbo]{Giulio Colombini} 
\ead{giulio.colombini2@unibo.it}
\author[gssi]{Nicola Guglielmi}
\ead{nicola.guglielmi@gssi.it}
\author[unibo,infnbo]{Armando Bazzani\corref{corresponding}}
\ead{armando.bazzani@unibo.it}

\cortext[corresponding]{To whom any correspondence should be addressed.}

\affiliation[unibo]{organization={Dipartimento di Fisica e Astronomia ``Augusto Righi''\\ Alma Mater Studiorum - Università di Bologna},
            addressline={Via Irnerio 46}, 
            city={Bologna},
            postcode={40126}, 
            country={Italia}}

\affiliation[infnbo]{organization={Istituto Nazionale di Fisica Nucleare, Sezione di Bologna},
            addressline={Viale Carlo Berti-Pichat 6/2}, 
            city={Bologna},
            postcode={40127}, 
            country={Italia}}
            
\affiliation[gssi]{organization={Gran Sasso Science Institute},
            addressline={Viale Francesco Crispi 7}, 
            city={L'Aquila},
            postcode={67100}, 
            country={Italia}}

\begin{highlights}
    \item Delay Differential Equations (DDEs) can model emergent states in directed dynamical networks.
   \item On a directed cycle of neurons a dynamical steady state emerges through a saddle-node bifurcation.
   \item Simulations reveal bidirectional mapping between DDE and network solutions.
   \item Both systems exhibit multiattractivity, but different attraction basins.
\end{highlights}
\begin{abstract}
While synchronized states, and the dynamical pathways through which they emerge, are often regarded as the paradigm to understand the dynamics of information spreading on undirected networks of nonlinear dynamical systems, when we consider directed network architectures, dynamical stationary states can arise.
To study this phenomenon we consider the simplest directed network, a single cycle, and excitable FitzHugh-Nagumo (FHN) neurons.
We show numerically that a stationary dynamical state emerges in the form of a self-sustained traveling wave, through a saddle-point bifurcation of limit cycles that does not destabilize the global fixed point of the system.
We then formulate an effective model for the dynamical steady state of the cycle in terms of a single-neuron Delay Differential Equation (DDE) featuring an explicitly delayed feedback, demonstrating numerically the possibility of mapping stationary solutions between the two models.
The DDE based model is shown to reproduce the entire bifurcation, which also in this case does not destabilize the global fixed point, even though global properties differ in general between the systems.
The discrete nature of the cycle graph is revealed as the origin of these coordinated states by the parametric analysis of solutions, and the DDE effective model is shown to preserve this feature accurately.
Finally, the scaling of the inter-site propagation times hints to a solitonic nature of the wave state in the limit of large chain size.
\end{abstract}


\end{frontmatter}



\section{\label{sec:intro}Introduction\protect}
The main focus of Complex networks theory has been the understanding of the topological structure of networks that represent interactions between nodes, corresponding to separate systems, while neglecting the dynamical properties of the nodes \cite{newman_networks_2018}. 
Nonetheless in most cases the network structure is inferred exactly via the effects of the interactions it represents on the global dynamics of the system,  making a joint study of structure and dynamics relevant in principle also for network reconstruction problems.
To try and bridge this gap, diffusion processes on networks have been considered to study the relationship between the spectral properties of the Laplacian matrix associated to a network and the corresponding dynamical properties of different random walks dynamics in the relaxation toward a stationary state \cite{schnakenberg_network_1976}. 
Whenever nonlinearity is introduced in the system, the typical emergent property that has been considered to stem from the interaction structure and node dynamics is synchronization, which has been characterized using the Master Stability Function framework \cite{arenas2008}, recently extended by \cite{bayani_transition_2024} to study the cascade of events that leads to the onset of global synchronization in network-coupled dynamical systems, highlighting how the network structure affects the dynamical path to macroscopic synchronized states.
In the latter case the most widely used model is an ensemble of $N$ identical dynamical systems coupled by a network structure
\begin{equation}
 \dot x_n=f(x_n)-J\sum_{m=1}^N \mathcal{L}_{nm} g(x_m)\qquad n=1,...,N
 \label{eq:chain:model_intro}
\end{equation}
where $x_n\in \mathbb{R}^d$ is the dynamical state of the $n$-th node, that evolves according to a local vector field $f(x)$ and an interaction term, specified through $\mathcal{L}$, a Laplacian matrix ($\sum_m \mathcal{L}_{nm}=0$) describing the network structure, a global real factor $J \geq 0$ modulating the coupling strength and an output function $g(x)$ which can describe the processing undergone by the coordinates of single systems before being fed into other nodes as input.
The synchronous state $x_m(t)=y(t)\,\, \forall\,m$ is an equilibrium state of the system (\ref{eq:chain:model_intro}) where $y(t)$ is an orbit of the non-interacting system for $J=0$.
The matrix $\mathcal{L}_{nm}$ introduces a self-feedback dissipative effect ($\mathcal{L}_{nn}\ne 0$) so that the interaction depends on the differences in the dynamical state between different nodes.
If the matrix $\mathcal{L}$ is symmetric, all the eigenvalues are positive real, and the synchronizability of a given dynamical systems ensemble is entirely specified by its Master Stability Function \cite{pecora_master_1998, arenas2008}, which only depends on the local features of the dynamics, and in particular on the maximum Lyapunov exponent of the tangent dynamics to the synchronized solution, up to a scaling of the coupling strength.
In particular, if the single node dynamics has a stable fixed point $y(t)=y_0$, the solution $x_n=y_0\,\forall\,n$ is a global stable solution of the system and we have a particular case of synchronized state corresponding to all nodes resting at the fixed point, and if the system $\dot y=f(y)$ has a periodic attractive solution, the equilibrium state corresponds to a synchronous oscillation of all the nodes with the same phase. 
In this framework the main problems are the study of the robustness of stable synchronized states \cite{buscarino2013robustness}, and the study of the relationships between the relaxation process, the spectral properties of the interaction matrix and the effect of stochastic perturbations \cite{song2022stochastic}.
\par\noindent From a statistical mechanics point of view, the symmetry of the $\mathcal{L}$ matrix corresponds to the detailed balance (DB) condition, i.e.~$\mathcal{L}_{ij} p_j = \mathcal{L}_{ji} p_i,\, \forall i, j \in \{1,\dots, N\}$ where $p_i$ is the stationary distribution of the random walk associated to $\mathcal{L}$ \cite{schnakenberg_network_1976}, so that a master equation on the associated network describes a diffusion process with a self-adjoint operator and a maximum entropy equilibrium state. 
Therefore, the fluctuations in a node's state spread across the network with a delay that depends on the local dynamics without the presence of any stationary probability current. 
The applications of complex network theory \cite{estrada2012structure} to physical systems like neuronal networks \cite{markov_weighted_2014}, epidemic spread \cite{pastor-satorras_epidemic_2015} and traffic dynamics \cite{schneider_unravelling_2013}, have repeatedly pointed out that Laplacian matrices associated to real networks do not satisfy the DB property, in most cases due to an intrinsically directed nature of the couplings.
In such cases diffusion processes on the networks relax towards a non-equilibrium stationary state (NESS), in which stationary probability currents exist.
The existence of such currents is strictly related to the cycle space of the graph, since there are as many independent currents as there are basis vectors in the cycle basis, a fact that reflects the stochastic irreversibility of the microscopic dynamics.

When one considers coupled dynamical systems on a graph, the emergence of stable global dynamical states due to the breaking of DB is also highly dependent on the features of the local dynamics.
When the single node dynamics has an attractive fixed point and the interaction structure is established through a Laplacian matrix, we have a constant global solution.
In the case of non-symmetric Laplacian matrices, simulations \cite{nagatani2002physics} suggest the existence of periodic stationary solutions, whose appearance can be related to phase transitions of the system \cite{hohenberg2015introduction} as the consequence of a bifurcation phenomenon in the phase space.
A relevant observation highlighted by numerical simulations is that the bifurcation phenomena can be explained by a low-dimensional system, even if the number of degrees of freedom is very large.
We conjecture that the existence of such bifurcations is related to the presence of probability currents in the stationary state of the associated diffusion process, that break the symmetry of the synchronized solution in networked dynamical systems, creating a traveling stationary wave in the network.
This is the case of dynamical neuronal networks, in which the existence of loops could induce feedback processes that create self-consistent stable periodic solutions. Due to the possible physiological role of loops as origin of feedback, these stationary solutions could be related to the emergence of memory mechanisms in brain structures.\par\noindent
In this article we address the problem of the existence of periodic stable stationary states for an ensemble of FitzHugh-Nagumo (FHN) neurons coupled in a directed cycle when the dynamics of each neuron has a single equilibrium point. 
This is the fundamental building block to understand how the graph geometry affects the global dynamics and we perform extensive numerical simulations to study the dependence of the solution on the model parameters.
While the synchronized solution of the system corresponds to all neurons lying near the stable equilibrium state of single node dynamics, when the coupling strength reaches a critical threshold, simulations highlight the rise, through a saddle-point bifurcation, of a stable periodic solution so that the global equilibrium state remains attractive.
To characterize the periodic solution we use a self-consistent approach based on a delay differential equation (DDE) whose periodic solution corresponds to the stationary solution of the neuronal network.
We numerically study the bifurcation phenomenon on the DDE to perform a quantitative comparison with the neuronal network dynamics.
The study of the existence of periodic solutions to DDEs has been considered \cite{yanchuk2019temporal} as modeling systems that exhibit temporal dissipative solitons. 
Due to the stiff character of the FHN dynamics, specific algorithms (namely \texttt{RADAR5} \cite{guglielmi1999order, guglielmi2001implementing, guglielmi2008computing}) have to be used to integrate the DDE.

\section{\label{sec:problem}The coupled FitzHugh-Nagumo oscillators model \protect}

We consider the case of interacting FHN neurons as a relevant example of nonlinear dynamical systems on a graph due to possible applications in neuroscience. The solution of dynamical neuronal networks could highlight some relevant phenomena that allow a better understanding of the dynamics of biological networks. 
In a general case, we consider an ensemble of interacting FHN neurons according to the equations
\begin{equation}
    \label{eq:chain:fhn_net}         
    \begin{aligned}
        \varepsilon\dot{u}_n &= u_n - \dfrac{u_n^3}{3} - v_n + J\sum_m\left (w_{nm}u_m - w_{mn}u_n\right)\\
        \dot{v}_n            &= u_n + a
    \end{aligned}   
\end{equation}
where $w_{nm}>0$, the link weights, define the interaction network and quantify the effect of the state of neuron $m$ on neuron $n$. 
We recall the $u_n$ represents the cell potential of the $n$-th neuron and $v_n$ a recovery variable, in principle representing several other internal variables \cite{izhikevich2007dynamical, fitzhugh196impulses}.
The parameter $J$ defines a global scaling of the coupling strength in the neuronal network, and $\varepsilon > 0$ is the time scale separation between the dynamics of the $u$ and the $v$.
This means that if the evolution of the $v$ variable takes a time $\bigo{1}$, the one of the $u$ variable takes a time $\bigo{\varepsilon}$.
To have biological significance, the parameter must be taken $\varepsilon \ll 1$, thus yielding a singularly perturbed problem \cite{kevorkian2012multiple}.
The real parameter $a$ determines the qualitative dynamics of the single neurons: for $|a|<1$ the fixed point \eqref{eq:chain:fixed_point} of the single-neuron dynamics is unstable and the system possesses a stable limit cycle, while for $|a| > 1$ the system undergoes a supercritical Andronov-Hopf bifurcation, the fixed point becomes attractive and the neuron is quiescent unless an external forcing is provided \cite{fitzhugh196impulses,izhikevich2007dynamical}.
The sum
$$
I_n^i(t)=J\sum_m w_{nm}u_m(t)
$$
is the input signal provided to neuron $n$ from the other neurons whereas 
$$
I_n^o(t)=J\sum_m w_{mn}u_n(t)=w_n u_n(t)
$$
is the total output signal from neuron $n$ that is introduced as a dissipative effect in its dynamics, where we have defined the symbol $w_n = \sum_m w_{mn}$ for the sum of the weights of the links exiting node $n$. 
Being interested in studying the role of fluctuations in stationary states, we assume the Laplacian condition for the weights
\begin{equation}
	w_{n} = \sum_m w_{mn} =\sum_{m} w_{nm}
  \label{eq:chain:balance}
\end{equation}
which means that for each node the sum of the weights of incoming and outgoing links is balanced, and that the global synchronized state $(u_n(t),v_n(t))=(u(t),v(t))$ is a solution of the single neuron dynamics. 
We choose the parameter $a$ so that the single node dynamics has a single stable fixed point and the synchronized solution reduces to the equilibrium position $(u^\ast,v^\ast)$ (see \eqref{eq:chain:fixed_point}).
We look for the existence of stationary attractive solutions $u_n^\infty(t)$ of the system \eqref{eq:chain:fhn_net}
\begin{equation}
\label{eq:chain:statsol}
u^\infty_n(t)=\lim_{T\to\infty} u_n(t+T)
\end{equation}
given a suitable initial condition. 
Averaging the dynamics over the network nodes, in order to perform a Mean Field Approximation (MFA), would provide
\begin{equation}
\label{eq:chain:meanfield}
\begin{aligned}
        \frac{\varepsilon}{N}\sum_{n = 1}^{N} \dot{u}_n &=\frac{1}{N}\sum_{n = 1}^{N} \left ( u_n - \dfrac{u_n^3}{3} - v_n\right )\\
        \frac{1}{N}\sum_{n = 1}^{N} \dot{v}_n &= \frac{1}{N}\sum_{n = 1}^{N} u_n + a
\end{aligned}
\end{equation} 
where the couplings disappear due to the Laplacian condition on the weights.
By introducing the average network activity variables
\[U(t)=\frac{1}{N}\sum_{n = 1}^{N} u_n(t),\quad V(t)=\frac{1}{N}\sum_{n = 1}^{N} v_n(t)\] and expressing the local variables in terms of fluctuations w.r.t.~the average, if the higher order fluctuation at each site can be neglected, one would recover the single neuron dynamics for the average network activity variables $U(t),\,V(t)$.
When one considers directed graphs ($w_{nm} \neq w_{mn}$), we will show numerically in Sect.~\ref{sec:numerics} that the MFA cannot describe the signal propagation observed in neuronal networks and the presence of feedback effects due to the loops can create new stationary global solutions that are not solutions of the averaged equation \eqref{eq:chain:meanfield}.

\begin{figure}
    \centering
    \resizebox{\linewidth}{!}{
\begin{tikzpicture}[
    sito/.style = {circle, minimum size=3pt, inner sep=0pt, fill=black},
    vuoto/.style = {minimum size=3pt, inner sep=0pt},
    etichetta/.style = {anchor=north}
]
\draw  (6, 0) arc(-90:90:0.5) -- (0, 1) arc(90: 270 : 0.5) -- cycle;
\foreach \x in {0,2,3,4,6}
    \node[sito] (sito-\x) at (\x, 0) {};

\node at (2.5, 0.5) {$J(u_{i-1} - u_i)$};

\node[vuoto] (sito-1) at (1, 0) {};
\node[vuoto] (sito-5) at (5, 0) {};
\node[etichetta] at (sito-0.south) {$1$};
\node[etichetta] at (sito-1.south) {$\mathellipsis$};
\node[etichetta] at (sito-2.south) {$i-1$};
\node[etichetta] at (sito-3.south) {$i$};
\node[etichetta] at (sito-4.south) {$i+1$};
\node[etichetta] at (sito-5.south) {$\mathellipsis$};
\node[etichetta] at (sito-6.south) {$N$};
\draw[bend left = 30, ->, > = latex, semithick, shorten <= 3pt, shorten >= 3pt] (sito-2) to (sito-3);
\end{tikzpicture}
}
    \caption{Schematic representation of the neuron loop.}
    \label{fig:neuron_ring}
\end{figure}
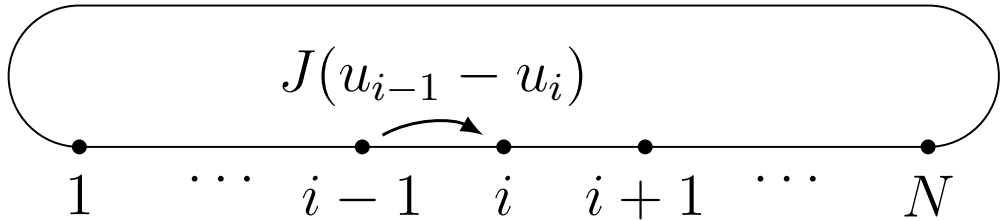

To study the existence of such solutions we consider a simplified model where $N$ FHN neurons interact by a directed loop and the general network-coupled system \eqref{eq:chain:fhn_net} reduces to
\begin{equation}
    \label{eq:chain:fhn_ring}         
    \begin{aligned}
        \varepsilon\dot{u}_n &= u_n - \dfrac{u_n^3}{3} - v_n + J \left(u_{n-1} - u_n\right)\\
        \dot{v}_n            &= u_n + a
    \end{aligned}   
\end{equation}
where the only free parameter is the excitatory coupling strength $J>0$ and we identify site $N+1$ and site $1$ imposing a periodic boundary condition, see Fig.~\ref{fig:neuron_ring} for a schematic representation. 
We choose the parameter value $|a| > 1$ so that the equilibrium solution of the FHN oscillator 
\begin{equation}\label{eq:chain:fixed_point}
u^\ast=-a\qquad v^\ast=\frac{a^3}{3} - a
\end{equation} 
is a stable equilibrium for the whole system (\ref{eq:chain:fhn_ring}) and the single-node FHN system does not admit any other solutions.
We recall that in the case $|a| < 1$ the single neuron exhibits a stable limit cycle, which for an undirected network implies the existence of a synchronized solution $u_n(t)=u(t)$ for all $n$, with all neurons moving synchronously on the limit cycle of the single FHN oscillator, whose stability depends on the strength of the coupling $J$ \cite{plotnikov_synchronization_2019}.
Indeed the single neuron dynamics undergoes a supercritical Andronov-Hopf bifurcation at $|a| = 1$ and a stable limit cycle appears, contextually to a loss of stability of the fixed point \cite{izhikevich2007dynamical, fitzhugh196impulses}.
In the considered case $|a| > 1$ the only global equilibrium is the fixed point \eqref{eq:chain:fixed_point} that is stable, so that the existence of a non-trivial stationary solution \eqref{eq:chain:statsol} has to be the result of a global bifurcation of different nature, that does not affect the fixed point stability.\par\noindent 
The variables $u_n$ and $v_n$ evolve respectively on a fast time scale and a slow time scale. 
The fast time scale $\varepsilon$ can be viewed as the reaction time scale of the neuron, i.e. the time needed to reach the peak of its action potential after receiving a triggering signal, whereas the slow time scale of order $\bigo{1}$ is the proportional to the time of relaxation to the single neuron equilibrium state.
In the physiologically relevant regime of $\varepsilon \ll 1$, the system (\ref{eq:chain:fhn_ring}) becomes numerically stiff so that the search for a global periodic solution requires the use of specific numerical integration schemes \cite{guglielmi2001implementing} (see also \ref{appa}).
Noticeably, the limit $\varepsilon\to 0$ is singular, due to the discontinuity determined by the jump between the two nullclines, and we lose the regularity of the solutions at $\varepsilon = 0$.
However, numerical simulations at very small $\varepsilon$ values suggest that the limit solution still exists despite losing continuity.
\par\noindent
We address the existence problem of stable stationary solutions for a feed-forward loop of neurons which do not change the stability of the fixed point \eqref{eq:chain:fixed_point} (i.e. the physiological resting state of the neuronal network) by looking for a self-consistent solution 
\begin{equation}
    \label{eq:chain:wave_ansatz}
    \begin{aligned}
    u_n(t) &= u\left(t - n\delta\right)\\
    v_n(t) &= v\left(t - n\delta \right),
    \end{aligned}
\end{equation}
that corresponds to a wave propagating in the direction of increasing site number. 
The time interval $\delta$ is the propagation time between two consecutive neurons and depends on the dynamics of the neurons themselves.
To obtain the relationship between $\delta$ and the other parameters, one should in principle study the advanced differential equation
\begin{equation}
    \label{eq:chain:fhn_ade}
    \begin{aligned}
        \varepsilon \dot{u} &= u - \frac{u^3}{3} - v + J \left(u(t + \delta) - u(t)\right)\\
        \dot{v} &= u + a,
    \end{aligned}
\end{equation}
with periodic boundary conditions $u(t+T)=u(t)$ and $v(t+T)=v(t)$, taking as period $T=N\delta$.
The bifurcations of \eqref{eq:chain:fhn_ade} correspond to the bifurcations of a traveling wave solution to the entire ring of excitable neurons, and the critical values of the parameters could in principle be obtained self-consistently by using a perturbation approach in the $\delta$ parameter.
However, from simulations (see Sect.~\ref{sec:numerics}) we observe that $\delta$ cannot be made arbitrarily small without destroying the wave solution, so that the condition $\delta\ll 1$ that would justify a perturbative approach is not satisfied.
The Cauchy problem for \eqref{eq:chain:fhn_ade} is ill-posed in general due to its advanced character. Letting 
\begin{equation}
\label{eq:chain:deviations_number}
  \tau=T-\delta=T\frac{N-1}{N},  
\end{equation}
the existence of a stationary periodic solution of \eqref{eq:chain:fhn_ring}, and equivalently of \eqref{eq:chain:fhn_ade}, corresponds to a solution of a Delay Differential Equation (DDE)
\begin{equation}
    \label{eq:chain:fhn_dde}
    \begin{aligned}
        \varepsilon \dot{u} &= u - \frac{u^3}{3} - v + J (u_\tau - u)\\
        \dot{v} &= u + a,
    \end{aligned}
\end{equation}
where $u_\tau = u(t - \tau)$ and $\tau$ is the positive delay defined in \eqref{eq:chain:deviations_number}. 
We shall see that the limit $N\to\infty$ corresponds to $T\to\infty$ and $\delta=T/N$ remains finite so that the stationary solution tends to a soliton wave in the system (\ref{eq:chain:fhn_ring}).\par\noindent
A stationary state of period $T$ for \eqref{eq:chain:fhn_ring} is a solution of the DDE with $\tau$ given by \eqref{eq:chain:deviations_number}, but, since no specific boundary condition is required in the solution of \eqref{eq:chain:fhn_dde}, we cope with the problem of the existence of periodic solutions of the DDE for a delay $\tau$ when the coupling parameter $J$ overcomes a critical threshold for a given $\epsilon$.
For $|a|>1$ the fixed point \eqref{eq:chain:fixed_point} is a stable solution of \eqref{eq:chain:fhn_dde}.
Using a heuristic argument, one observes that for $J=0$ the dynamics is dissipative in the whole phase space, but the FHN system is able to create an orbit that follows the stable branch of the nullcline and relaxes to the fixed point, when the resting state is perturbed in a suitable way. 
If the average dissipation is small (i.e. $|a|-1\ll 1$) and the parameter $J$ is sufficiently large, the delayed term $J(u_\tau-u)$ can induce a bifurcation in phase space acting as continuous perturbation, thus creating a stable and an unstable periodic orbit.
If $T-\tau=\delta$ is weakly dependent on the period, the study of this bifurcation phenomenon in the DDE varying $\tau$ implies the existence of a stationary state for the neuron loop \eqref{eq:chain:fhn_ring} when $T/\delta=N$ is integer.
From the previous formulation of the problem, both the dynamical parameters of the single system $a$, $\varepsilon$ and the coupling strength $J$ will be the same across the two systems, whereas the relationship between the delay time $\tau$ and $N$ is given by \eqref{eq:chain:deviations_number}.
The dependence of the periodic solution on the coupling parameter $J$ must be the same for both cases.\par\noindent
The correspondence between the periodic solutions of the systems \eqref{eq:chain:fhn_ring} and \eqref{eq:chain:fhn_dde} only refers to the solution existence, but their stability and the measure of eventual attraction basins remains to be established.\par\noindent
We finally remark that the existence of a self-consistent wave solution is a consequence of the discrete character of the system and we shall see that we have solutions with a finite period only for finite $N$.

\subsection{Stability problem of the stationary solution}
The stability problem for the self-consistent periodic solution $u_n(t)=u(t-\delta n),\,v_n(t)=v(t - \delta n)$ requires to study the linearized system
\begin{equation}
    \label{eq:chain:fhn_ringlin}         
    \begin{aligned}
        \varepsilon\delta \dot{u}_n &= \left (1 - u^2(t-Tn/N)\right )\delta u_n - \delta v_n + J \left(\delta u_{n-1} - \delta u_n\right)\\
        \delta \dot{v}_n            &= \delta u_n,
    \end{aligned}   
\end{equation}
where $(\delta u_n,\delta v_n)$ are small perturbations.
Due to the periodic dependence on $t$, with period $T$, one can consider the Poincaré matrix of the system and study its spectral properties.
The stability property requires that for any choice of the perturbation the linearized system has all eigenvalues with a negative real part, so that the critical values of the parameters correspond to the existence of an imaginary eigenvalue.
If one considers \eqref{eq:chain:fhn_dde} linearized at the periodic solution $u(t),\,v(t)$, the equation for the periodic orbit's stability is obtained, in the form
\begin{equation}
    \label{eq:chain:fhn_ddelin}
    \begin{aligned}
        \varepsilon \delta \dot{u} &= \left (1 - u^2(t)\right )\delta u - \delta v + J (\delta u_\tau - \delta u)\\
        \delta \dot{v} &= \delta u,
    \end{aligned}
\end{equation}
and the periodicity of the solution to \eqref{eq:chain:fhn_dde} allows us to look for a solution in Floquet form 
\begin{displaymath}
	\delta u=e^{\lambda t}\delta \bar u(t),\quad \delta v=e^{\lambda t}\delta \bar v(t),
\end{displaymath}
where $\lambda\in\mathbb{C}$ is the Floquet exponent, corresponding to the Lyapunov exponent of the Poincar\'e map, and $\delta \bar u(t),\, \delta\bar v(t)$ are periodic with period $T$.
Inserting this form in \eqref{eq:chain:fhn_ddelin} we get
\begin{equation}
    \label{eq:chain:fhn_ddelin2}
    \begin{aligned}
        \varepsilon \delta \dot{\bar u} &= -\lambda \delta \bar u +\left (1 - u^2(t)\right )\delta \bar u - \delta \bar v + J (e^{-\lambda \tau}\delta \bar u_\tau - \delta \bar u)\\
        \delta \dot{\bar v} &= -\lambda \delta \bar v+\delta \bar u.
    \end{aligned}
\end{equation}
These solutions correspond to the eigenvectors of the Poincar\'e matrix computed at a given section $t_s\in[0,T]$ with $\lambda$ their associated eigenvalue.
By definition we have
\begin{displaymath}
\delta \bar u(t-\tau)=\delta \bar u(t+T/N),\quad \delta \bar v(t-\tau)=\delta \bar v(t+T/N)
\end{displaymath}
and if we set 
\begin{equation}
    \begin{aligned}
       \delta u_n(t)&=e^{\lambda t} \delta \bar u(t-Tn/N) \\
       \delta v_n(t)&=e^{\lambda t} \delta \bar v(t-Tn/N)
    \end{aligned}
\label{eq:solpert}
\end{equation}
we get a solution for the system (\ref{eq:chain:fhn_ringlin}). 
Therefore, if \eqref{eq:chain:fhn_ddelin2} admits a periodic solution $\delta \bar u(t)$ with $\mathfrak{R}(\lambda)>0$, we have instability for the self-consistent wave solution of the initial system \eqref{eq:chain:fhn_ring}.
Since the stability of the origin of the Poincar\'e matrix implies linear stability for the self-consistent wave, the study of the periodic solutions of \eqref{eq:chain:fhn_ddelin2} when $\lambda$ varies allows to solve the stability problem, in an approach reminiscent of the Master Stability Function framework, developed for the synchronization of dynamical systems \cite{pecora_master_1998}.

\section{\label{sec:numerics}Numerical simulations results\protect}
We have studied the relationship between the system \eqref{eq:chain:fhn_ring} and the DDE \eqref{eq:chain:fhn_dde} by performing extensive numerical simulations. 
The purpose of this parametric analysis is to study the existence of periodic solutions for the network \eqref{eq:chain:fhn_ring}, and the dependence of their properties on the system size, coupling strength and time scale separation.
An analytical study would be highly not trivial due to the nonlinear character of the system.

\subsection{\label{subsec:chain}Chain solution dependence on the parameter values}

\begin{figure*}
    \begin{subfigure}{0.5\linewidth}
    \centering
    \includegraphics[width=\linewidth]{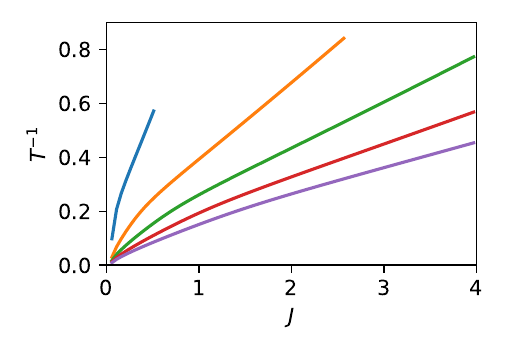}
    \caption{}\label{fig:v_vs_J:pana}
    \end{subfigure}
    \begin{subfigure}{0.5\linewidth}
    \centering
    \includegraphics[width=\linewidth]{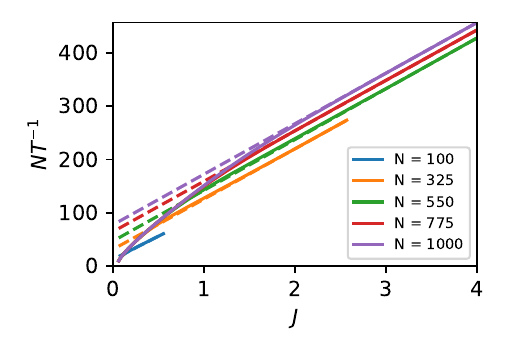}
    \caption{}\label{fig:v_vs_J:panb}
    \end{subfigure}
    \caption{Panel (a): self-consistent periodic wave speed as a function of $J$ for several values of the chain size $N$. 
	     From top to bottom $N = 100$ (blue)$,\, 325$ (orange) $,\, 550$ (green)$,\, 725$ (red)$,\, 1000$ (purple).
	     Panel (b): the same quantity multiplied by the chain size $N$.
	     From bottom to top $N = 100$ (blue)$,\, 325$ (orange) $,\, 550$ (green)$,\, 725$ (red)$,\, 1000$ (purple).
	     The dashed lines mark the linear fits performed on the final $25\%$ of the curve for each value of chain size.
             In both panels the interruption on the right end of the two lines $N = 100$ and $N = 325$ marks the disappearance of the wave solution.}
    \label{fig:v_vs_J}
\end{figure*}

To explore the dependence of the wave solution on the parameters, in particular on the number of neurons $N$ and on the coupling strength $J$, we study how the wave propagation speed varies as a function of the coupling strength $J$ for several sizes $N$.
Since no explicit length scale can be identified in the model, we identify the wave speed as the inverse of the period, i.e. we fix without loss of generality the length of the chain to unit. 
The simulation results are collected in Fig.~\ref{fig:v_vs_J}.
Each of the selected system sizes displays the same value for the lower threshold of the coupling, below which no stable wave solution exists.
Repeating simulations with several other values of $N$, the critical coupling value of $J \approx 0.05$ appears to remain the same independently of the selected system size.
Above the critical value the wave speed begins to grow in function of the coupling, settling later into a roughly linear growth regime for each chain size value.
The function appears to be concave.
An interesting feature of the system is the disappearance of the wave solution for large values of $J$. 
This is clearly visible in the $N = 100$ and $N = 325$ plots of Fig.~\ref{fig:v_vs_J}, but has been observed to happen, at larger values of $J$, for the other values of $N$ as well.
By looking at the velocity in units of sites per unit time in Figure \ref{fig:v_vs_J:panb}, we observe that at larger values of $J$ the growth approaches a linear trend, of slope apparently common among the various chain size values, despite the presence of an offset which does not allow for a full superimposition.
Performing a linear fit on the trend for each value of chain size we obtain the coefficients reported in Table \ref{tab:linearfits}.
\begin{table}[h]
\centering
\begin{tabular}{@{}S[table-format=4, table-column-width = 3em]| *2{S[table-format=2.2+-1.2, table-column-width = 6em]} @{}}
	\toprule
	   {$N$} & {slope} & {intercept} \\
	\midrule
     	100 	 &   86.67+-0.08 & 12.87+-0.04 \\
     	325 	 &   94.18+-0.03 & 31.55+-0.06 \\
     	550 	 &   95.15+-0.02 & 47.06+-0.06 \\
     	725 	 &   94.49+-0.02 & 64.76+-0.04 \\
     	1000 	 &   94.75+-0.01 & 77.64+-0.02 \\
	\bottomrule
    \end{tabular}
    \caption{Values for the slopes and intercepts of the linear fits to the plots in Fig.~\ref{fig:v_vs_J}b.
    The average slope value is $93 \pm 3$, where the uncertainty is attributed as the standard deviation of the sample.}
    \label{tab:linearfits}
\end{table}
\begin{figure}
    \centering
    \includegraphics[width=\linewidth]{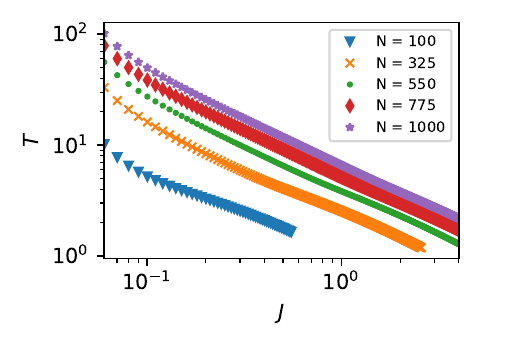}
    \caption{Dependence of the period $T$ of the ring solution on the coupling $J$ for several values of $N$ in a log-log scale. The interruption of the $N = 100$ and $N = 325$ series on the right hand side of the plot indicates disappearance of the wave solution. The values of $N$ with the corresponding symbols are 100 blue triangles, 325 orange crosses, 550 green dots, 725 red diamonds and 1000 purple stars.}
    \label{fig:T_vs_J}
\end{figure}
\begin{figure}
    \centering
    \includegraphics[width=\linewidth]{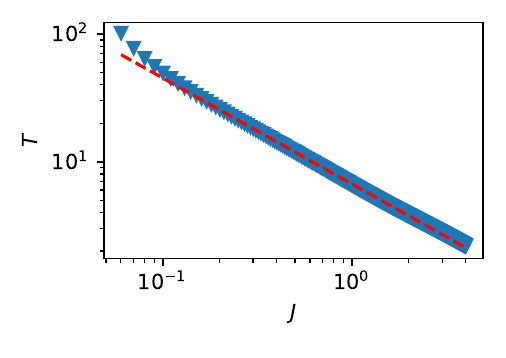}
    \caption{Dependence of the period $T$ on the coupling $J$ for a chain of $N = 1000$ sites. The dashed red line is a power law $T \propto J^{-k}$ fit to the right-most tail ($J \geq 2\times10^{-1}$), with an exponent $k = -0.8296 \pm 0.0007$.}
    \label{fig:fit_T_vs_J}
\end{figure}
In Fig.~\ref{fig:T_vs_J} we plot the dependence of the period $T$ of the ring wave solution on the coupling strength $J$ in a log-log scale, where we can see that as $N$ grows larger and for the greater values of $J$ the trend generally adapts to a power law.
A power law fit T $\propto J^{-k}$ to the tail of the plot for $N = 1000$, reproduced in Fig.~\ref{fig:fit_T_vs_J}, yields an exponent $k = -0.8296 \pm 0.0007$ pointing out the reasonable existence of some finer underlying trend with respect to the apparently linear growth observed in Fig.~\ref{fig:v_vs_J}.
To interpret these findings we propose the following picture.
The lower coupling threshold is a consequence of the nonlinear nature of the system: $J$ plays the role of magnification/attenuation factor for the preceding neuron signal, and as a consequence of the choice of the coupling increments the linear dissipation by an extra term $Ju_n$.
In this context, the lower critical value of $J$ can be interpreted as the minimal amount by which a spike from a FHN neuron can be scaled, while remaining able to elicit a spike in an identical unit, considering also the extra dissipation term $Ju_n$ introduced in both units by the Laplacian coupling.
In light of this interpretation, also the linear trends in Fig.~\ref{fig:v_vs_J} can be explained.
Indeed if we assume that existence of the wave solution for a given $J$ requires that input from each preceding neuron rises to a sufficiently high level in a sufficiently short time to produce a spike in its following neuron, doubling the value of $J$ will intuitively have the effect of halving the time it takes to reach said critical value.
Since the time scale of the fast dynamics is $\varepsilon$ we expect that analogous phenomena can take place when the latter is varied.
From this discussion, and in the absence of scaling w.r.t.\ $N$ in the coupling, it is apparent that the existence of self-sustained site to site transmission of the neuronal pulse is a matter that entirely depends on the satisfaction of local requirements at each of the links, but does not depend on any global coordination effect in the solution.
On the other hand, the disappearance of the wave solution for high values of the coupling can be justified by considering the increasing trend of the wave speed w.r.t.\ $J$, such that at some point the wave is too fast, and travels all the way around the ring, trying to excite neurons which are still in a refractory phase, i.e.\ too far from the fixed point to be displaced across the central nullcline branch by the spike of an identical neuron, so that a spike cannot be elicited any longer in a self-sustained way, and the wave disappears.
This picture could also explain how longer chains can attain larger speeds before the solution disappears w.r.t.~shorter ones, even though at the same coupling value they always admit slower waves in comparison, the proposed reason being that due to the lack of a natural spatial scale in the system, in chains with more sites the wave can reach higher speeds without colliding with refractory neurons which cannot be excited yet.
A relevant feature of the discussed self-consistent periodic wave is that throughout its existence the stable fixed point of the system remains such, as it would be expected with a directed Laplacian coupling, so that the observed phase transition cannot be regarded as an Andronov-Hopf bifurcation of the fixed point, but rather a saddle-point bifurcation of limit cycles.
Through this bifurcation a pair of periodic orbits appears in the system, the attractive one, which we observe, and a repulsive one.

\subsection{\label{subsec:dde}Delay Differential Equation solution dependence on the parameter values}

\begin{figure}[!htb]
    \centering
    \includegraphics[width=\linewidth]{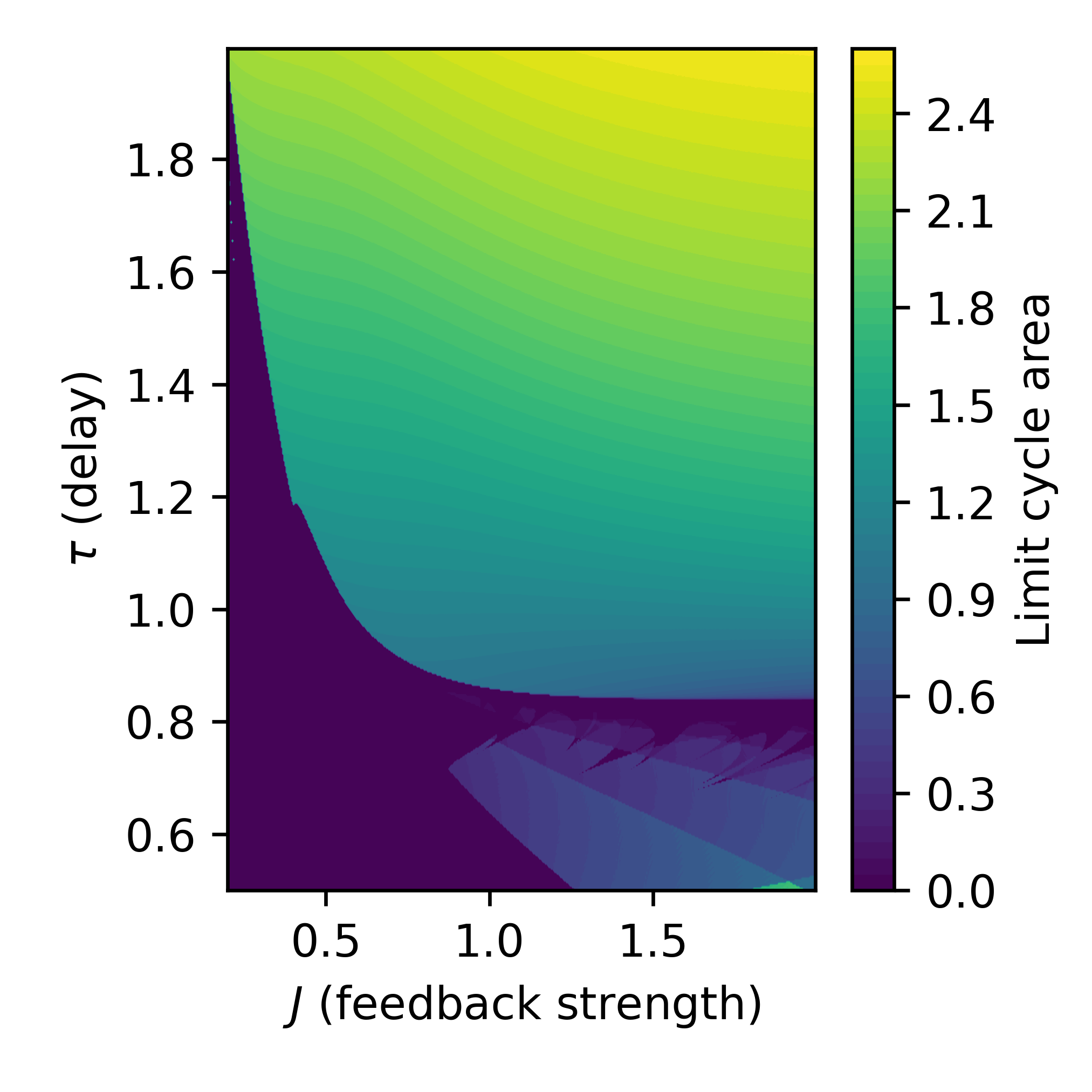}
    \caption{Phase diagram for the Delay Differential Equation \protect{\eqref{eq:chain:fhn_dde}}. The colour scale indicates the area of the limit cycle. In the lower right corner of the heatmap, the colour seems to suggest the existence of a limit cycle, but is in fact an artifact due to long transients before relaxation to the fixed point. The parametrization used for the initial spike is the one of \protect{\eqref{eq:init_not_normalised}} with $A = (\sqrt{2 \pi} \sigma)^{-1}$, $\sigma = 10^{-1}$.}
    \label{fig:DDE_phase_diagram}
\end{figure}

\begin{figure}[!htb]
    \centering
    \includegraphics[width=\linewidth]{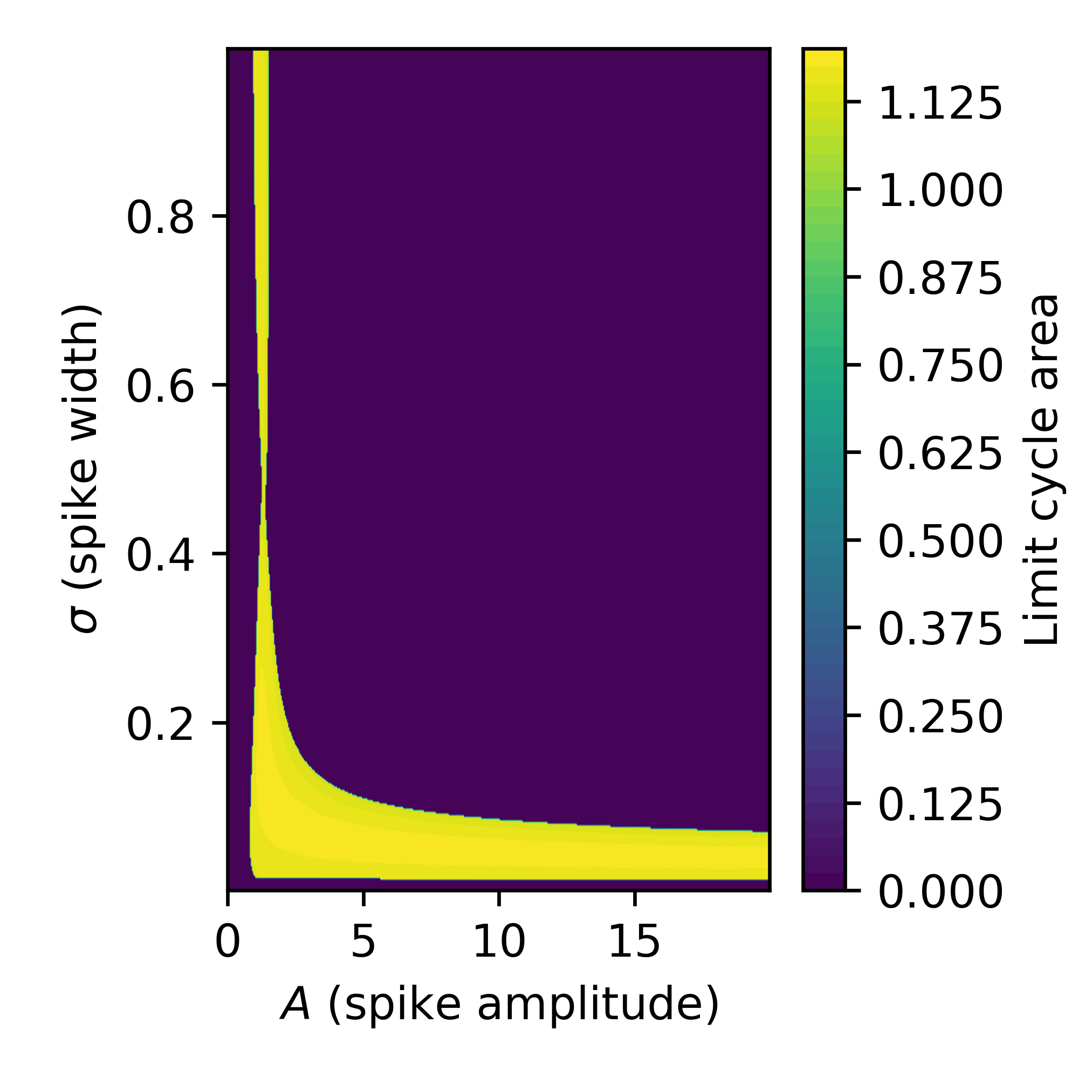}
    \caption{Area of the limit cycle in function of the amplitude $A$ and width $\sigma$ of the initial pulse for equation \eqref{eq:chain:fhn_dde} with $J = 1$ and $\tau = 1$.}
    \label{fig:DDE_attractivity}
\end{figure}

In this paragraph we study the properties of the limit cycle solution to the DDE \eqref{eq:chain:fhn_dde}, versus the choice of parameters of the system $J$ and $\tau$.
In particular we are interested in the existence of a sharp transition between a quiescent neuron, and a fully active and continuously spiking one.
To study these properties we perform repeated simulations of the system, varying $J \in \left[0.2, 2\right]$ and $\tau \in \left[0.5, 2\right]$ and calculate the area of the limit cycle that appears in the system, when it exists.
In delayed systems the initial condition must be supplied as a function over the interval $\left[-\tau, 0\right]$ so to have a defined value for the delay term during the first $\tau$ of integration.
We choose to initialize the system at it fixed point, providing a gaussian stimulus in the fast variable $u$
\begin{equation}\label{eq:init_not_normalised}
\begin{aligned}
    \varphi_u(t) &= -a + A e^{-\left(t - \tau/2\right)^2/2\sigma^2}\\
    \varphi_v(t) &= -a + a^3/3
\end{aligned}
\end{equation}
To study the dependence of the solution on on the parameters $J$ and $\tau$ we normalize the gaussian pulse by setting $A = (\sqrt{2 \pi} \sigma)^{-1}$.
The value of choice for the width parameter has been set to $\sigma = 10^{-1}$.
The results of this simulation are collected in Fig.~\ref{fig:DDE_phase_diagram}.
For any fixed $J$ we observe that there exists a minimal value of $\tau$ below which the limit cycle is not created, and above which the cycle sharply appears, and a similar picture appears if we fix $\tau$, highlighting the existence of a minimal critical $J$.
Moreover we observe an increasing trend of the limit cycle area in function of the delay for a fixed $J$, and an analogous increasing trend versus $J$ for a fixed $\tau$.
The properties of the solution for delayed systems can in principle depend strongly on the initial condition, due to the infinite dimensionality of the problem.
To assess how much the properties of the initial spike influence the solution we remove the normalization condition, so that we can act independently on its width $\sigma$ and amplitude $A$.
As in the previous case, for each simulation we measure the value of the limit cycle area, when one is created.
In this case, we also perform extensive simulations, the results of which are collected in Fig.~\ref{fig:DDE_attractivity}.
The emerging picture is that for both $A$ and $\sigma$ there exist optimal values, for which the value of the other parameter becomes unimportant.
Our interpretation of the optimal value of $A$ is that it is the value for which the system starts out optimally displaced across the central nullcline, far enough from the fixed point to elicit a spike, but not too far, where the dissipative component of the dynamics would be dominant. 
Indeed, since the selected value for the delay is $\tau = 1$, as $\sigma$ grows closer to $1$ we give the system a nearly constant input, but we are still able to reach the limit cycle.
We also observe the existence of a minimal $\sigma$ value, which we interpret reasonably as the minimal duration that a pulse must have to generate a response in the system.
On the other hand, for the optimal $\sigma$ range around $0.1$, the proposed explanation is that in this interval we are feeding the system with pulses of a similar duration to those that will appear in the limit cycle, in a sense putting it in closer proximity to it w.r.t. other initial condition choices. 
In this context as well, we observe a minimal $A$ value below which, despite the optimality of the pulse shape, the input is too low to cause any activity in the system.
In addition to these findings, we observe that whenever the limit cycle exists, there is hardly any dependence of the actual value of the area on the system parameters.
As in the chain, the stable fixed point of the system $(u^*, v^*)$ is not destabilized by the appearance of the limit cycle, so that an Andronov-Hopf bifurcation can also be ruled out in this case.

\subsection{\label{subsec:correspondence}Correspondence of the solutions}

\begin{figure*}
    \centering
    \includegraphics[width = \linewidth]{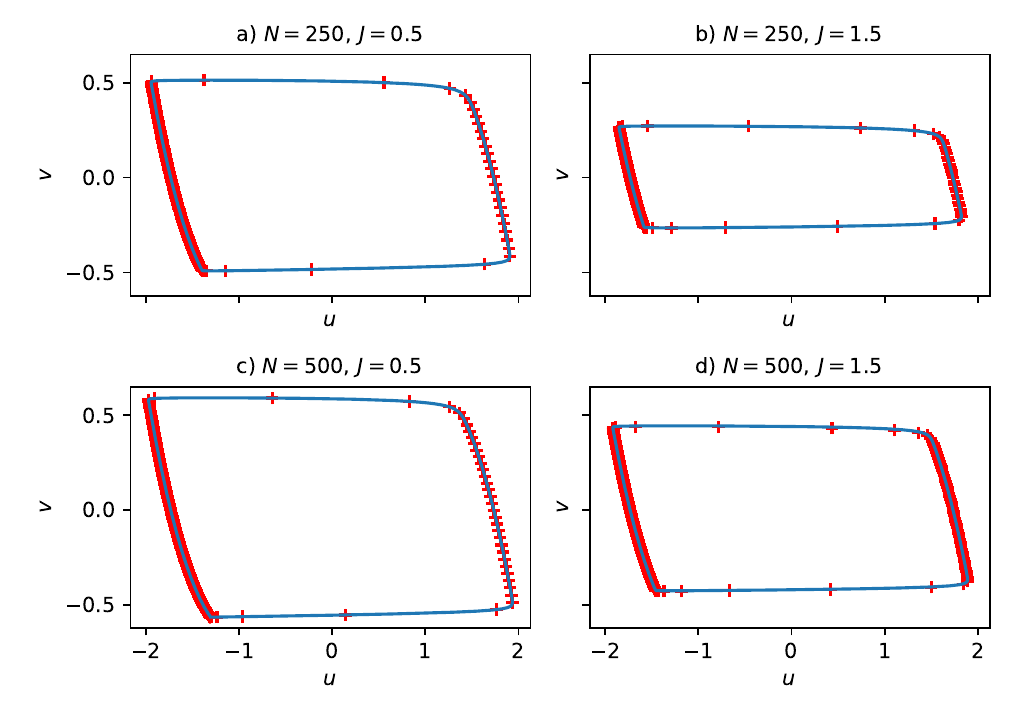}
    \caption{Examples of the limit cycles of the delay equation (continuous blue line), and the chain sites (red crosses). Notice how the cycle area decreases with larger couplings and increases with the number of neurons in the chain.}
    \label{fig:lc_table}
\end{figure*}

We compare numerically the self-consistent periodic solutions for the loop of neurons with those of the single delay differential equation \eqref{eq:chain:fhn_dde}. 
To do so we fix a number of neurons $N$ and check with a simulation for the existence of a stable periodic solution of the system \eqref{eq:chain:fhn_ring}. 
If it exists we calculate its period $T$ by measuring the time interval between two consecutive crossings of a Poincar\'e section of the single site dynamics. 
Then the relation \eqref{eq:chain:deviations_number} determines the delay time $\tau$ to be inserted in eq.\ \eqref{eq:chain:fhn_dde} to get the corresponding periodic solution. 
In Fig.~\ref{fig:lc_table} we show some examples of comparison between the continuous stable periodic solution of eq.\ \eqref{eq:chain:fhn_dde} and the self-consistent periodic wave solution of the neuron ring defined by the neuron states at a given time.

We observe that the numerical solutions of the delay differential equation interpolate perfectly the periodic solutions of the FHN neuron loop when the parameters are varied, and that due to the stiffness of the problem most of the chain sites are at any given time on the nullclines of the system.
We can also infer some qualitative aspects of the limit cycle dependence on the system parameters.
Namely, raising the value of the coupling $J$ makes the cycle thinner along the vertical axis, this is easily noticeable e.g. comparing insets a) and b) or c) and d) from Fig.~\ref{fig:lc_table}.
This observation suggests that in the context of the chain model the wave solution speed is a non-decreasing function of the coupling $J$, causing the system to close the cycle in shorter and shorter times as $J$ is increased.
Since changing the system size $N$ at fixed $J$ amounts to changing the delay $\tau$ in the DDE, it is easy to match the a) $\to$ c) or b) $\to$ d) trend with the one observed for fixed $J$ in Fig.~\ref{fig:DDE_phase_diagram}.
Conversely, the apparently contrasting decreasing trend of the area from a) $\to$ b) or from c) $\to$ d) can be justified by remembering that when $J$ is changed at fixed system size in the chain, the equivalent delay of the DDE is influenced as well, through the dependence of $T$ on $J$, therefore in the phase diagram of Fig.~\ref{fig:DDE_phase_diagram} the equivalent displacement to an increase of $J$ in the chain is a diagonal motion towards higher coupling values and smaller delays.
We use the same Gaussian external stimulus \eqref{eq:init_not_normalised}, with $A = 12.5,\, \sigma = 10^{-1}$, for the delay differential equation and we let the system relax to the attractive periodic solution, when it exists. 
In both cases the periodic solution is a stable attractive state of the dynamics with a defined stability basin.
The numerical integration of the delay differential equation has been performed using the RADAR5 \cite{guglielmi1999order, guglielmi2001implementing, guglielmi2008computing} implicit integration algorithm, which is an extension of the 3-stage Radau IIA implicit Runge-Kutta method \cite{hairer2010solvingII} to a broad class of (possibly implicit) delay differential equations.
In Table \ref{tab:gap_comparison} we report the numerical values of the two sides of \eqref{eq:chain:deviations_number} obtained for the values of the coupling $J$ and $N$ of Fig.~\ref{fig:lc_table}, for a fixed value of the time scale separation parameter $\varepsilon = 10^{-2}$.
We observe an agreement between the r.h.s.~and the l.h.s.~of \eqref{eq:chain:deviations_number} in the considered cases.
The likely source of the small differences observed in some cases reasonably lies in the different attractivity of the solutions between the DDE and the unidirectional chain, i.e. in a different rate of convergence to the periodic solution in the two cases.
\begin{table}[h]
\centering
    \begin{tabular}{@{}S[table-format=3, table-column-width = 2em]S[table-format=1.1, table-column-width = 2em]|S[table-format=1.3, table-column-width = 4em] S[table-format=1.3, table-column-width = 4em]@{}}
	\toprule
	   {$N$} & {$J$} & {$T - \tau$} & {$T/N$} \\
	\midrule
     250  & 0.5 &   0.0134  &  0.0139 \\
     250  & 1.5 &   0.0057  &  0.0060 \\
     500  & 0.5 &   0.0125  &  0.0119 \\
     500  & 1.5 &   0.0053  &  0.0053 \\
	\bottomrule
    \end{tabular}
    \caption{Values of the phase difference between two consecutive chain sites and the corresponding gap between the DDE solution period and the delay. 
             The value of $\tau$ used to compute the second quantity is the equivalent delay as calculated via \protect{\eqref{eq:chain:deviations_number}}. All reported digits have been found to be stable across three orders of magnitude of tolerance of the integrator ($10^{-9}$ to $10^{-12}$).}
    \label{tab:gap_comparison}
\end{table}

\begin{figure}
    \centering
    \includegraphics[width = \linewidth]{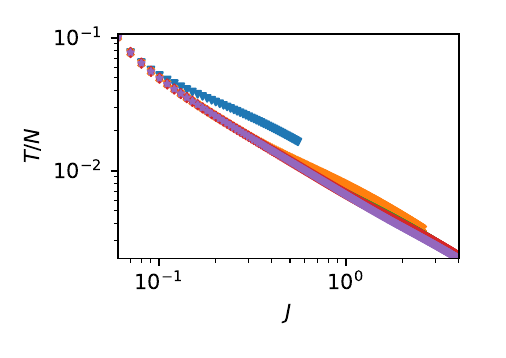}
    \caption{Dependence of the $T/N$ site-to-site propagation delay on the coupling strength $J$ for several values of chain size $N$. The values of $N$ with the corresponding symbols are 100 blue triangles, 325 orange crosses, 550 green dots, 725 red diamonds and 1000 purple stars.}
    \label{fig:ToN_vs_J}
\end{figure}

On both sides of \eqref{eq:chain:deviations_number} there is an implicit dependence, via $T$, on the coupling strength $J$, therefore we can check their scaling versus the coupling.
In Fig.~\ref{fig:ToN_vs_J} we show the dependence on $J$ of $T/N$, this quantity corresponds to the parameter $\delta$ of Sec.~\ref{sec:problem}, and amounts to the time it takes for the wave to travel from a site to the following one.
We observe that all system sizes superimpose on each other for most of the plot, while for $N = 100$ and $N = 325$ we observe a deviation from the power law behaviour immediately before the disappearance of the wave solution. This can be interpreted as a slowdown of the wave due to the fact that its front is beginning to encounter neurons earlier and earlier into their refractory phase.

Furthermore, the superimposition becomes clearer and sharper with growing $N$ for fixed $J$, as the finite size effects require larger and larger coupling values to become noticeable.
This observation suggests a well defined $N\to\infty$ limit for the inter-site time $T/N$, corresponding to a self-sustained traveling perturbation also in the limit of an infinitely long chain.

\begin{figure}
    \centering
    \includegraphics[width=\linewidth]{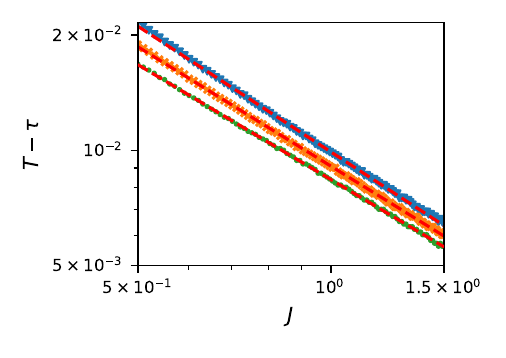}
    \caption{Dependence of the $T - \tau$ gap on the coupling strength $J$ for several values of the delay $\tau$. The values of $\tau$ with the corresponding symbols are $1$ blue triangles, $1.5$ orange crosses and $2$ green dots. The dashed red lines are power law $T - \tau \propto J^{-k}$ fits. From top to bottom the exponents are for $\tau = 1$, $k=1.092 \pm 0.003$, for $\tau = 1.5$, $k = 1.038 \pm 0.002$ and for $\tau = 2$, $k=0.999 \pm 0.002$.}
    \label{fig:fit_Ttau_vs_J_dde}
\end{figure}

We proceed similarly for the Delay Differential Equation, this time fixing $\tau$ while varying $J$. 
In Fig.~\ref{fig:fit_Ttau_vs_J_dde} we plot $T - \tau$, the quantity analogous to $T/N$ for the delay system, which in this case can be interpreted as the advance of the delay w.r.t.~the final period of the solution. 
Also in this case we observe a power law behaviour, which upon fitting in the form $T - \tau \propto J^{-k}$ reveals exponents $\approx -1$, exact values with uncertainties are reported in the caption of Fig.~\ref{fig:fit_Ttau_vs_J_dde}.

\subsection{\label{subsec:stiffness}Dependence on $\varepsilon$}
\begin{figure}
    \centering
    \includegraphics[width=\linewidth]{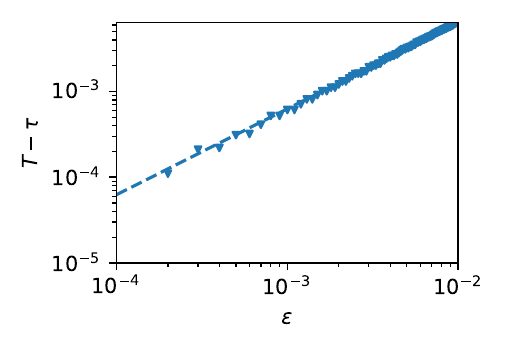}
    \caption{$T - \tau$ gap dependence on the value on the stiffness parameter $\varepsilon$, for the Delay differential equation. The coupling and the delay were fixed at values $J = 1.5$, $\tau = 1$. The dashed line is a power law $T - \tau \propto \varepsilon^k$ fit, yielding an exponent $k = 1.002 \pm 0.008$.}
    \label{fig:Ttau_vs_epsilon}
\end{figure}

In the whole of the preceding part of this work, we kept $\varepsilon = 10^{-2}$ fixed, but in principle the properties of the self-consistent wave/limit cycle solution should depend on it, since it represents the reaction time scale of the system, setting the time of $\mathcal{O}(\varepsilon)$ that it takes for the system to jump from a nullcline to the other, and therefore to react whenever a sufficiently strong and sharp input is provided.
From the discussion of Section \ref{subsec:chain} we expect that decreasing $\varepsilon$ in the chain system \eqref{eq:chain:fhn_ring} will have a similar effect on the global properties of the solution as increasing $J$.
On a single node the two scalings remain distinguishable, as decreasing $\varepsilon$ forces the single node dynamics to adhere more closely to the nullclines, while the $J$ scaling has no such effect.
Indeed, we observed in simulations that depending on the size of the system there exists an inferior critical $\varepsilon$, below which the wave solution disappears in the same way as it does for too large $J$, i.e. by becoming too fast and colliding with neurons that are still too refractory.
In agreement with the picture proposed for $J$, the critical $\varepsilon$ decreases when the size of the system is increased, as larger rings make it easier to accommodate faster waves.
This behaviour is strikingly similar to what is observed in traffic models such as \cite{nagatani1998modified}, where albeit originating from an explicit delay term, a too little reaction time causes the wave solution to destabilize and disappear.
Studying the DDE, instead, we observe no disappearance of the solution for small $\varepsilon \approx 10^{-12}$.
This property makes it easier to study the dependence, e.g. of $T$ on $\varepsilon$, and we can study the l.h.s.\ of \eqref{eq:chain:deviations_number} $T - \tau$ at fixed $\tau$ and $J$ by varying $\varepsilon$.
We do so by performing simulations, reported in Fig.~\ref{fig:Ttau_vs_epsilon}.
Consistently with what we expect, we observe a power law behaviour $T - \tau \propto \varepsilon^k$ with a positive exponent $k = 1.002 \pm0.008$ in function of $\varepsilon$.

\section{\label{sec:conclusion}Conclusion\protect}
The purpose of this work is to investigate how breaking the detailed balance condition in a network of nonlinear dynamical systems can give rise to self-sustained steady dynamical states.
The choice of nonlinear spiking neurons has been motivated by their relevance in models of neuroscientific interest, while the focus on a simple loop structure is motivated by it being the simplest essentially directed structure, its natural role as a source of feedback in larger and more realistic networks, and the parallel existing between cycles and independent probability currents in NESS, in diffusive dynamics.
By means of extensive and accurate simulations, which were performed with methods suited to the intrinsic stiffness of the system at hand, we were able to find evidence that the self-consistent periodic wave that appears in such models does not arise as a consequence of a Hopf bifurcation of the system's fixed point, which remains stable and that it is a phenomenon without a natural length scale, but rather through a saddle-point bifurcation of limit cycles, introducing multiattractivity in the system.
More specifically, for the trend of the wave speed an explanation has been proposed in terms of the system nonlinearity, and proof has been found that the wave formation is a consequence of the balanced interplay between the discrete nature of the system and the finite reaction time scale $\varepsilon$.
We show evidence supporting these arguments by examining the mechanism through which the wave disappears for large coupling values and that an increase in the coupling is to an extent equivalent to a reduction of the reaction time scale.
Furthermore, the analysis of the inter-node propagation time in the directed periodic chain hints to a well-defined traveling pulse state also in the limit of an infinitely long chain.
This kind of interplay is similar to what is observed e.g. in traffic models \cite{nagatani1998modified}, pointing to a degree of universality.
In addition, we are able to formulate an equivalence between the wave solution to the directed cycle of FHN units and a limit cycle in a single neuron Delay Differential Equation with an explicitly delayed feedback term.
The mapping of systems and solutions is bidirectional so that given the parameters of the DDE or of the chain, and the solution period, it is possible to calculate the parameters of the equivalent system.
However, attraction basins are observed to be different between the chain and the DDE, as is expected since they are global features of the system.
Similarly, the dissipation rate of solutions approaching the self-consistent periodic solution/limit cycle is found to differ in the two systems, the network being much quicker in relaxing to the self-consistent solution than the DDE to the limit cycle.

The natural continuation of the work hereby presented would be to move towards large networks of a more complex structure, which would be relevant from several points of view.
Firstly, the presence of loops, possibly supporting self-sustained activity, could play a role in shaping the information spreading patterns on a network \cite{hens2019spatiotemporal}, for example by suppressing or enhancing signals depending on their timing with respect to local activity levels.
This perspective could be further expanded in the context of information theory, which would require an attentive discretization of the dynamics of nonlinear neurons such as FHN units, so to preserve the features that we have identified as key in the creation of self-sustained activity.
Finally, in the context of biologically inspired neural networks, the correlation between the presence of cyclic structures in a neural network, and the sensitivity of the machine learning algorithms relying on it to periodic content in the input could be investigated.
This could be done in the context of neurally inspired reservoir learning \cite{albeattie2024criticality}, by purposefully looking at the cycle structure of reservoir graphs that yield different efficiencies when analyzing periodic signals, or in a more interesting context by studying whether a training performed on the reservoir weights themselves tends to alter the loop structure of the network.

\section*{Declaration of competing interest}
The authors declare that they have no known competing financial interest or personal relationships that could have appeared to influence the work reported in this paper.
\section*{CRediT authorship contribution statement}
\noindent\textbf{Giulio Colombini:} Conceptualization, Methodology, Software, Writing - original draft, Writing - review and editing, Visualization. 

\noindent\textbf{Nicola Guglielmi:} Methodology, Software, Supervision.

\noindent\textbf{Armando Bazzani:} Conceptualization, Supervision, Writing - original draft, Writing - review and editing. 

\section*{Data availability}
No experimental data was used for the research reported in the present article.
\appendix
\section{Numerical details}
\label{appa}
We gather in this appendix some details concerning technical aspects of the simulations performed for this article.
The simulations of the chain \eqref{eq:chain:fhn_ring} were performed through the Radau method routines \cite{hairer2010solvingII} provided by the Scipy package \cite{2020scipy}, with the initial perturbation being an interpolation on the chain sites of the gaussian pulse in the fast variable \eqref{eq:init_not_normalised} parameterised as follows
\begin{equation}\label{eq:init_chain}
\begin{aligned}
    u_n(0) &= -a + A e^{-(n - \mu_N)^2/2\sigma_N^2}\\
    v_n(0) &= a - a^3/3
\end{aligned}
\end{equation}
with $\mu_N = N/2,\, \sigma_N = 10^{-2} N,\, A = 2$.
The integrator was set to operate with a relative and an absolute tolerance of $10^{-4}$, and an initial step size of $10^{-6}$ (note that the algorithm makes use of a variable step size based on local error estimation).
All the speed and period values reported in Figures\ \ref{fig:v_vs_J}, \ref{fig:T_vs_J}, \ref{fig:fit_T_vs_J} and \ref{fig:ToN_vs_J} were calculated as follows.
First the maximum of the wave profile, corresponding to the rising wave front was located for each integration timestep. Then the location values were mapped to a monotonically nondecreasing time series by accounting for the periodicity of the system, and thus adding an $N$ for each revolution around the ring.
Finally, the monotonic time series trend has been fitted with a line versus time, so to obtain the wave speed as the angular coefficient. To perform the linear regression Scipy \cite{2020scipy} routines were used.
All the aforementioned simulations had a duration of $25$ time units, which has been deemed sufficient to let the transients relax in the chain systems.

The simulations for the Delay Differential Equation \eqref{eq:chain:fhn_dde} were all performed with the RADAR5 \cite{guglielmi1999order, guglielmi2001implementing, guglielmi2008computing} integrator.
Both the relative and absolute tolerance of the integrator were set at $10^{-12}$, and the initial step size to $10^{-6}$. In Figures \ref{fig:DDE_phase_diagram} and \ref{fig:DDE_attractivity} the system was simulated for a time of $30 \tau$, the area of the limit cycle being calculated from the average of the cycles present in the final $30\%$ of the trajectory.
The detection and separation of cycles has been performed by means of a Poincar\'e section in the form of a vertical line through the fixed point.
The calculation of areas has been performed via the well-known Surveyor's formula.
In both Fig.~\ref{fig:fit_Ttau_vs_J_dde} and Fig.~\ref{fig:Ttau_vs_epsilon}, the simulation duration was of $100 \tau$ for each instance of the system, the reason for the longer runs being the attempt to mitigate the effect of transients in finer measurements.
Periods were calculated here as well by means of Poincar\'e sections.

For Fig.~\ref{fig:lc_table} there was again a need for mitigation of transients so that a total duration of $80$ time units was selected for both the chain and the DDE simulations.
The isolation of the DDE limit cycle has been performed via Poincar\'e section.
Initial pulses were of the form \eqref{eq:init_chain} with $A = 2.5$ for the chain and of the form \eqref{eq:init_not_normalised} with $A = 12.5$ for the DDE.
The reason for the larger pulse in the DDE is that from our studies the delayed system appears much more dissipative than the chain, so that even if a stable limit cycle exists, a too weak initial pulse could be unable to push the system onto it.
All linear fits in the article have been performed via the \texttt{linregress} routine present in the Scipy package \cite{2020scipy}, and all power laws have been fitted using the \texttt{curve\_fit} routine from the same package.
\bibliographystyle{elsarticle-num}
\bibliography{chain}

\end{document}